\documentclass[letterpaper, 10 pt, conference]{ieeeconf}
\IEEEoverridecommandlockouts
\usepackage{cite}
\usepackage{amsmath,amssymb,amsfonts}
\usepackage{algorithmic}
\usepackage{graphicx}
\usepackage{textcomp}
\usepackage{xcolor}

\newtheorem{lemma}{Lemma}
\newtheorem{theorem}{Theorem}

\allowdisplaybreaks

\title{\LARGE \bf
Hybrid PHD-PMB Trajectory Smoothing Using Backward Simulation
}

\author{Yuxuan Xia$^{1}$, {\'A}ngel F. Garc{\'i}a-Fern{\'a}ndez$^{2}$, and Lennart Svensson$^{3}$% <-this % stops a space
\vspace{-3mm}
\thanks{This work was partially supported by the Wallenberg AI, Autonomous Systems and Software Program (WASP) funded by the Knut and Alice Wallenberg Foundation.}% <-this % stops a space
\thanks{$^{1}$Yuxuan Xia is with Zenseact, Gothenburg, Sweden, and the Department of Electrical Engineering, Link{\"o}ping University, Link{\"o}ping, Sweden. {\tt\small yuxuan.xia@liu.se}}%
\thanks{$^{2}${\'A}ngel~F.~Garc{\'i}a-Fern{\'a}ndez is with the Department of Electrical Engineering and Electronics, University of Liverpool, Liverpool, United Kingdom, and ARIES Research Centre, Universidad Antonio de Nebrija, Madrid, Spain. {\tt\small angel.garcia-fernandez@liverpool.ac.uk}}%
\thanks{$^{3}$Lennart Svensson is with the Department of Electrical Engineering, Chalmers University of Technology, Gothenburg, Sweden. {\tt\small lennart.svensson@chalmers.se}}
}

\begin{document}

\maketitle
\thispagestyle{empty}
\pagestyle{empty}

\begin{abstract}
The probability hypothesis density (PHD) and Poisson multi-Bernoulli (PMB) filters are two popular set-type multi-object filters. Motivated by the fact that the multi-object filtering density after each update step in the PHD filter is a PMB without approximation, in this paper we present a multi-object smoother involving PHD forward filtering and PMB backward smoothing. This is achieved by first running the PHD filtering recursion in the forward pass and extracting the PMB filtering densities after each update step before the Poisson Point Process approximation, which is inherent in the PHD filter update. Then in the backward pass we apply backward simulation for sets of trajectories to the extracted PMB filtering densities. We call the resulting multi-object smoother hybrid PHD-PMB trajectory smoother. Notably, the hybrid PHD-PMB trajectory smoother can provide smoothed trajectory estimates for the PHD filter without labeling or tagging, which is not possible for existing PHD smoothers. Also, compared to the trajectory PHD filter, which can only estimate alive trajectories, the hybrid PHD-PMB trajectory smoother enables the estimation of the set of all trajectories. Simulation results demonstrate that the hybrid PHD-PMB trajectory smoother outperforms the PHD filter in terms of both state and cardinality estimates, and the trajectory PHD filter in terms of false detections.
\end{abstract}

% \begin{IEEEkeywords}
% Multi-object tracking, smoothing, probability hypothesis density, Poisson multi-Bernoulli, backward simulation, sets of trajectories.
% \end{IEEEkeywords}

\section{Introduction}

Multi-object filtering concerns the estimation of the number of objects and their current states based on a sequence of noisy measurements \cite{meyer2018message,streit2021analytic}. This problem is compounded by missed detections, clutter, and data association uncertainties. To address the multi-object filtering problem in a computationally efficient way, the probability hypothesis density (PHD) filter recursively propagates the first-order moment approximation of the multi-object posterior density, which is also known as the PHD \cite{phdpoint}. By doing so, the PHD filter operates on the single-object space and, consequently, avoids the data association problem arising from multiple objects. Due to its computational efficiency, the PHD filter has been widely used in many applications \cite{phdpoint2,ristic2011note,phdextended2,phdextended3,lundquist2010road,gao2021phd}.

The estimation of multi-object states by also using measurements at later time steps is called multi-object smoothing. The PHD smoother involves first a forward PHD filtering recursion and then a backward smoothing recursion \cite{nadarajah2011multitarget,mahler2012forward}; the latter is performed by propagating the PHD of the multi-object smoothing density backward in time. Particle filter implementations of the PHD smoother were provided in \cite{nadarajah2011multitarget,mahler2012forward,nagappa2011fast,georgescu2013particle,feng2016adaptive} and the closed-form Gaussian implementations were presented in \cite{vo2011closed,he2016improved}. However, it has been reported in \cite{mahler2012forward,nagappa2011fast,he2016improved} that, compared to the PHD filter, the PHD smoother reduces the localization error, but it does not necessarily improve the cardinality estimate. In particular, the PHD smoother suffers from premature object deaths \cite{nagappa2011fast} by reporting missed detections at earlier time steps when objects disappear, with a lag equal to the smoothing lag.

% It was reported in \cite{mahler2012forward,nagappa2011fast,he2016improved} that the PHD smoother can correct state estimation errors, but that it does not necessarily improve the cardinality estimates. In particular, objects in the PHD smoother suffer from premature deaths with a lag equal to the smoothing lag \cite{nagappa2011fast}.

A common drawback of the PHD filter and smoother is that they do not maintain track continuity, and thus, cannot provide object trajectory estimates. A simple track building procedure for the PHD filter is by adding labels/tags to the object states or PHD components, and track continuity can be established by linking the object states or PHD components with the same label/tag over time \cite{panta2009data,lu2017labeled}. These methods may work well in some scenarios but can also result in track switches, and even missed and false detections, see \cite{garcia2019trajectory}. 

A mathematically rigorous way to build trajectories for the PHD filter without using any labels/tags is by considering sets of trajectories \cite{garcia2019multiple}. The resulting filter is called the trajectory PHD filter, which directly provides object trajectory estimates by recursively propagating a Poisson multi-object density on the space of sets of trajectories \cite{garcia2019trajectory}. Furthermore, smoothed trajectories can be obtained in forward filtering by considering smoothing-while-filtering. However, the trajectory PHD filter is mainly used for estimating trajectories of alive objects, and it cannot properly estimate the set of all trajectories, which also includes trajectories of dead objects.

The possibility of generating sets of smoothed trajectories for both alive and dead objects using a sequence of (unlabeled) multi-object filtering densities has been explored in \cite{xia2022multiple}. In particular, this work showed that we can sample sets of trajectories using backward simulation \cite{lindsten2013backward} even when the forward multi-object filter is an unlabeled filter that does not explicitly maintain track continuity. The presented multi-trajectory smoothing solution has been applied to the Poisson multi-Bernoulli (PMB) filters \cite{pmbmpoint,variational}, and the resulting multi-trajectory smoother has achieved state-of-the-art multi-trajectory estimation performance. However, there are no similar methods for the PHD filters\footnote{Backward simulation was first applied to the particle-based PHD filter in \cite{georgescu2013particle} to generate smoothed trajectories with labels, but the algorithm developed in \cite{georgescu2013particle} can only track up to two objects.}.

% Therefore, it would be interesting to investigate how to apply backward simulation for sets of trajectories \cite{xia2022multiple} to the PHD filter. To do so, it is important to understand the connection between the PHD and PMB filters.

% The PHD filter considers a Poisson multi-object density, in which the cardinality of the set is Poisson distributed and, for each cardinality, its elements are independent and identically distributed (i.i.d.). In PHD filtering, the predicted multi-object density is still Poisson without approximation, but the updated multi-object density is of the form PMB \cite{pmbmpoint}, which consists of a Poisson point process (PPP) for missed detected objects and a multi-Bernoulli (MB) process for detected objects. The Bayesian filtering recursion is achieved by finding the Poisson approximation of the PMB density that matches its PHD. The PHD filter can also be considered as a special implementation of the PMB filter, where all the Bernoulli components in the MB are approximated as being Poisson (also called recycling) after each update step \cite{recycle}.

In this paper, we present a hybrid PHD-PMB trajectory smoother, which involves PHD forward filtering and PMB backward smoothing. In PHD filtering, the predicted multi-object density is of the form Poisson Point Process (PPP) without approximation, but the updated multi-object density is of the form PMB \cite{pmbmpoint}. The Bayesian filtering recursion is achieved by finding the Poisson approximation of the PMB density that matches its PHD. The key difference between our proposed hybrid PHD-PMB trajectory smoother and all the existing PHD smoothers \cite{nadarajah2011multitarget,mahler2012forward,nagappa2011fast,feng2016adaptive,vo2011closed,he2016improved,georgescu2013particle} is that we perform backward smoothing on the PMB multi-object densities in the PHD filtering recursion before the Poisson approximation. Considering that the PMB representation of the multi-object posterior is an inherent intermediate result that we obtain for free in the PHD update step
(as we will show in Section \ref{sec_hybrid_phd_pmb}), it makes sense to perform smoothing using this more informative multi-object posterior representation.

The hybrid PHD-PMB trajectory smoother leverages the simplicity of PHD filtering \cite{phdpoint2} and the ability to generate smoothed trajectories using backward simulation \cite{xia2020backward,xia2022multiple}. Importantly, the hybrid PHD-PMB trajectory smoother makes it possible to generate smoothed trajectories for all the objects that have existed in a theoretically sound manner without using any labels/tags. We compare the proposed hybrid PHD-PMB trajectory smoother to the PHD \cite{phdpoint2} and trajectory PHD \cite{garcia2019trajectory} filters in a simulation study, and the results demonstrate that the proposed method has superior multi-object state and trajectory estimation performance. The results also show that the hybrid PHD-PMB trajectory smoother is more robust to the premature object death problem \cite{mahler2012forward,nagappa2011fast}. 

The rest of the paper is organized as follows. In Section~II, we introduce the background. The problem formulation is given in Section~III. We present the filtering and smoothing recursions of the hybrid PHD-PMB trajectory smoother in Section IV. The results are presented in Section~V, and the conclusions are drawn in Section VI.

\section{Background}

In this section, we introduce the state variables of interest, densities on sets of objects and sets of trajectories, and the multi-object models. The set of finite subsets of a generic space $D$ is denoted by ${\cal F}(D)$, and the cardinality of a set $A\in {\cal F}(D)$ is $|A|$. The sequence of ordered positive integers $(\alpha,\alpha+1,\dots,\gamma-1,\gamma)$ is $\alpha:\gamma$. We use $\uplus$ to denote a union of sets that are mutually disjoint and $\langle f,g \rangle$ to denote the inner product $\int f(x)g(x) dx$. In addition, we use $\delta_x(\cdot)$ and $\delta_x[\cdot]$ to represent the Dirac and Kronecker delta functions, respectively, centered at $x$.

\subsection{State Variables}

The single-object state is described by a vector $x \in \mathbb{R}^{n_x}$. A trajectory is represented as a variable $X = (t,x^{1:\nu})$ where $t$ is the initial time step of the trajectory, $\nu$ is its length, and $x^{1:\nu} = (x^1,\dots,x^\nu)$ denotes a finite sequence of length $\nu$ that contains the object states at time steps $t:t+\nu-1$. For two time steps $\alpha$ and $\gamma$, $\alpha \leq \gamma$, a trajectory $(t,x^{1:\nu})$ in the time interval $\alpha:\gamma$ existing from time step $t$ to $t+\nu-1$ satisfies that $\alpha \leq t \leq t+\nu-1 \leq \gamma$, and the variable $(t,\nu)$ hence belongs to the set $I_{(\alpha,\gamma)} = \{(t,\nu):\alpha \leq t \leq \gamma~\text{and}~ 1\leq \nu \leq \gamma - t +1\}$. A single trajectory in time interval $\alpha:\gamma$ therefore belongs to the space $T_{(\alpha,\gamma)} = \uplus_{(t,\nu)\in I_{(\alpha,\gamma)}}\{t\}\times \mathbb{R}^{\nu n_x}$ \cite{garcia2019multiple}. 

A set ${\bf x} \in {\cal F}(\mathbb{R}^{n_x})$ of single-object states is a finite subset of $\mathbb{R}^{n_x}$, and a set ${\bf X}_{\alpha:\gamma} \in {\cal F}(T_{(\alpha,\gamma)})$ of trajectories is a finite subset of $T_{(\alpha,\gamma)}$. The subset of trajectories in ${\bf X}_{\alpha:\gamma}$ that were alive at time step $\eta$ where $\alpha \leq \eta \leq \gamma$ is denoted by 
\begin{equation}
  {\bf X}_{\alpha:\gamma}^{\eta} = \left\{ \left(t,x^{1:\nu} \right)\in {\bf X}_{\alpha:\gamma}: t \leq \eta \leq t+\nu-1 \right\}.
\end{equation}
Given a single-object trajectory $X = (t,x^{1:\nu})$, the set of object states at time step $k$ is 
\begin{equation}
  \tau^k(X) = \begin{cases}
    \left\{x^{k+1-t}\right\}, & t \leq k \leq t + \nu - 1 \\
    \emptyset, & \text{otherwise} 
  \end{cases}
\end{equation}
and given a set ${\bf X}_{\alpha:\gamma}$ of trajectories, the set of object states at time step $k$ is $\tau^k({\bf X}_{\alpha:\gamma}) = \bigcup_{X\in{\bf X}_{\alpha:\gamma}}\tau^k(X)$. 

\subsection{Densities on Sets of Objects}

Given a real-valued function $\pi(\cdot)$ on the space ${\cal F}(\mathbb{R}^{n_x})$ of finite sets of objects, its set integral is \cite{mahler2007statistical}
\begin{equation}\label{eq_set_integral}
  \int \pi({\mathbf{x}}) \delta {\mathbf{x}} = \pi(\emptyset) + \sum_{n=1}^{\infty}\frac{1}{n!}\int \pi\left(\left\{x_1,\dots,x_n\right\}\right) d (x_{1},\cdots, x_{n}).
\end{equation}
A function $\pi(\cdot)$ on the space ${\cal F}(\mathbb{R}^{n_x})$ is a multi-object density if $\pi(\cdot) \geq 0$ and its set integral is one. 

A PMB density is of the form \cite{pmbmpoint}
\begin{align}
    f({\bf x}) &= \sum_{\mathbf{y}\uplus \mathbf{w} = {\bf x}}f^{p}(\mathbf{y})f^{mb}(\mathbf{w}),\label{eq_pmb}\\
    f^{p}({\bf x}) &= e^{-\left\langle \lambda,1 \right\rangle} \prod_{x \in \mathbf{x}} \lambda(x),\label{eq_ppp}\\
    f^{mb}({\bf x}) &= \sum_{\uplus_{l=1}^{n}{\bf x}^l = {\bf x}}\prod_{i=1}^{n}f^i\left({\bf x}^i\right),\label{eq_mb}\\
    f^i({\bf x}) &= \begin{cases}
    1 - r^i & {\bf x} = \emptyset \\
    r^ip^i(x) & {\bf x} = \{x\} \\
    0 & \text{otherwise}.
    \end{cases}\label{eq_ber}
\end{align}
The PMB in \eqref{eq_pmb} is the union of two independent sets: a PPP with density \eqref{eq_ppp}, parameterized by Poisson intensity $\lambda(\cdot)$, and an MB with density \eqref{eq_mb}, where the $i$-th Bernoulli component has density \eqref{eq_ber} with existence probability $r^i$ and single-object density $p^i(\cdot)$.

\subsection{Densities on Sets of Trajectories}

Given a real-valued function $\pi(\cdot)$ on the single trajectory space $T_{(\alpha,\gamma)}$, its integral is \cite{garcia2019multiple}
\begin{equation}
  \int \pi(X) d X = \sum_{(t,\nu)\in I_{(\alpha,\gamma)}} \int \pi\left(t,x^{1:\nu}\right) d x^{1:\nu},
\end{equation}
which goes through all possible start times, lengths and object states of trajectory $X \in T_{(\alpha,\gamma)}$.
A function $\pi(\cdot)$ on the space ${\cal F}(T_{(\alpha,\gamma)})$ is a multi-trajectory density if $\pi(\cdot) \geq 0$ and its set integral is one. The trajectory PPP and trajectory MB densities are defined similar to their counterparts in \eqref{eq_ppp} and \eqref{eq_mb}, respectively. Given two single-object trajectories $X = (t,x^{1:\nu})$ and $Y = (t^\prime,y^{1:\nu^\prime})$, the trajectory Dirac delta function is defined as 
\begin{equation}
  \delta_{Y}(X) = \delta_{t^\prime}[t]\delta_{\nu^\prime}[\nu]\delta_{y^{1:\nu^\prime}}\left(x^{1:\nu}\right),
\end{equation}
and the multi-trajectory Dirac delta function centred at ${\bf Y}$ is defined as \cite[Sec. 11.3.4.3]{mahler2007statistical}
\small
\begin{equation}
  \delta_{{\bf Y}}({\bf X}) = \begin{cases}
    0 & |{\bf X}| \neq |{\bf Y}|,\\
    1 & {\bf X} = {\bf Y} = \emptyset,\\
    \sum_{\sigma\in\Gamma_n}\prod_{i=1}^n\delta_{Y_{\sigma_i}}(X_i) & \begin{cases}
      {\bf X} = \{X_i\}_{i=1}^n \\
      {\bf Y} = \{Y_i\}_{i=1}^n.
    \end{cases}
  \end{cases}
\end{equation}
\normalsize

Given a set ${\bf x}_k$ of object states at time step $k$, the set of trajectories in the time interval $k:k$ is 
\begin{equation}
  {\bf X}_{k:k} = \left\{X = \left(k,x^{1:1}\right): x^1 \in {\bf x}_k\right\},
\end{equation}
where trajectory $X=(t,x^{1:\nu})\in {\bf X}_{k:k}$ has start time $t=k$ and length $\nu=1$. Therefore, it holds that the multi-trajectory density $\pi({\bf X}_{k:k})$, which is zero for trajectories outside the interval $k:k$, takes the same value as the multi-object state density $f(\tau^{k}({\bf X}_{k:k}))$.

\subsection{Multi-Object Models}\label{sec_model}

We consider the standard multi-object models \cite{mahler2007statistical}. For a given multi-object state $\mathbf{x}_k$ at time $k$, each object state $x \in \mathbf{x}$ is either detected with probability $p^D(x)$ and generates one measurement $z$ with density $\ell(z|x)$, or missed detected with probability $1-p^D(x)$. The set $\mathbf{z}_k$ of measurements at time step $k$ is the union of the object-generated measurements and PPP clutter with intensity $\lambda^C(\cdot)$. 

Given the current multi-object state ${\bf x}_k$, each object $x \in {\bf x}_k$ survives with probability $p^S(x)$, and moves to a new state with a Markovian transition density $g(\cdot|x)$, or dies with probability $1-p^S(\cdot)$. The multi-object state at the next time step is the union of the surviving objects and new objects, which are born independently of the rest. The set of newborn objects is a PPP with intensity $\lambda^B(\cdot)$.

\subsection{Multi-Trajectory Dynamic Model}\label{sec_tra_predict_model}

Performing backward smoothing on multi-trajectory density requires a dynamic model for the set of all trajectories that have existed up to the current time step. Given a set ${\bf X}_{1:k}$ of all trajectories in the time interval $1:k$, each trajectory $X = (t,x^{1:\nu}) \in {\bf X}_{1:k}$ ``survives'' with probability one, $p^S(X) = 1$, and moves to a new state according to \cite{garcia2019multiple}
\begin{align}
  \label{eq_single_trajectory_transition}
  g^{k+1}\left(t^\prime,y^{1:\nu^\prime} | X\right) &= \left|\tau^{k}(X)\right| \left[ \left(1-p^S\left(x^\nu\right)\right)\delta_X\left(t^\prime,y^{1:\nu^\prime}\right)\right.\nonumber\\
  &+ \left. p^S\left(x^\nu\right) g\left(y^{\nu^\prime} | x^{\nu}\right)\delta_X\left(t^\prime,y^{1:\nu^\prime-1}\right) \right]\nonumber\\
  &+ \left(1 - \left|\tau^{k}(X)\right|\right)\delta_X\left(t^\prime,y^{1:\nu^\prime}\right).
\end{align}
That is, if the object trajectory $X$ has died before time step $k$, the trajectory remains unaltered with probability one. If trajectory $X$ exists at time step $k$, it remains unaltered with probability $1-p^S(x^\nu)$, or the new final object state $y^{\nu^\prime}$ is generated according to the single-object transition density with probability $p^S(x^\nu)$. The set ${\bf X}_{1:k+1}$ of trajectories in the time interval $1:k+1$ is the union of ``surviving'' trajectories and newborn trajectories, where each new trajectory $(t,x^{1:\nu})$ has a deterministic start time $t=k+1$, length $\nu=1$, and its multi-object state is a PPP with intensity $\lambda^B(\cdot)$. The trajectory PPP birth at time step $k+1$ has intensity
\begin{equation}
    \lambda^B_{k+1}\left(t,x^{1:\nu}\right) = \delta_{k+1}[t]\delta_1[\nu]\lambda^B\left(x^\nu\right).
\end{equation}

\section{Problem Formulation}

In backward smoothing for sets of trajectories, we aim to compute the multi-trajectory posterior density in a given time interval using a sequence of multi-object filtering densities and the multi-trajectory dynamic model.

We denote the multi-object state density at time step $k^\prime \in \{k,k+1\}$ and the multi-trajectory density in time interval $\alpha:\gamma$, both conditioned on the sequence of sets of measurements ${\bf z}_{1:k} = ({\bf z}_1,\dots,{\bf z}_k)$ up to and including time step $k$, by $f_{k^\prime|k}(\cdot)$ and $\pi_{\alpha:\gamma|k}(\cdot)$, respectively. In forward filtering, the Bayes filter propagates the multi-object posterior density of $\mathbf{x}_k$ in time via the prediction and update steps:
\begin{align}
    f_{k|k-1}({\bf x}) &= \int g({\bf x}|{\bf x}^\prime) f_{k-1|k-1}({\bf x}^\prime) \delta {\bf x}^\prime,\label{eq_predict}\\
    f_{k|k}({\bf x}) &= \frac{\ell({\bf z}_k|{\bf x})f_{k|k-1}({\bf x})}{\int \ell({\bf z}_k|{\bf x})f_{k|k-1}({\bf x}) \delta {\bf x}},\label{eq_update}
\end{align}
where $g(\cdot|{\bf x})$ is the multi-object transition density and $\ell({\bf z}_k|\cdot)$ is the measurement likelihood. Given the multi-object models described in Section \ref{sec_model}, the predicted and posterior densities on the current set of object states are PMB mixtures (PMBMs) \cite{pmbmpoint}. A PMB is a special case of PMBM with a single MB.

The multi-trajectory density $\pi_{1:K|K}(\mathbf{X}_{1:K})$ can be obtained by applying the following multi-trajectory smoothing equation recursively backwards in time \cite[Theorem 2]{xia2022multiple}:
\begin{equation}
    \label{eq_smoothing}
    \pi_{k:K|K}({\bf X}_{k:K}) = \frac{\pi_{k:k+1|k}({\bf X}_{k:k+1})\pi_{k+1:K|K}({\bf X}_{k+1:K})}{f_{k+1|k}(\tau^{k+1}({\bf X}_{k+1:k+1}))},
\end{equation}
where the predicted multi-trajectory density for the multi-trajectory dynamic model in Section \ref{sec_tra_predict_model} is given by \cite[Theorem 1]{xia2022multiple}
\small
\begin{align}\label{eq_multistep1}
    &\pi_{k:k+1|k}({\bf X}_{k:k+1}) = \prod_{\left(k,x^{1:\nu}\right)\in{\bf X}_{k:k+1}^{k}} \Bigg[ \left( 1+p^S\left(x^\nu\right) \left(\delta_{2}[\nu]-1\right) \right) \nonumber\\
    &~~~\times  \delta_{2}[\nu] g \left( x^{\nu} | x^{\nu-1}\right) p^S\left(x^{\nu-1}\right) \Bigg]\pi_{k:k|k}({\bf X}_{k:k}) \nonumber\\ &~~~\times  e^{-\left\langle \lambda_{k+1}^B,1 \right\rangle} \prod_{\left(k+1,x^{1:1}\right) \in \mathbf{X}_{k+1:k+1}} \lambda^B_{k+1}(k+1,x^{1:1}).
\end{align}
\normalsize
It can be observed that the predicted multi-trajectory density can be evaluated by multiplying the multi-trajectory density $\pi_{k:k|k}({\bf X}_{k:k})$, which takes the same value as $f_{k|k}(\tau^k({\bf X}_{k:k}))$, $1-p^S(\cdot)$ for trajectories that end at time step $k$, $g(\cdot|\cdot)p^S(\cdot)$ for trajectories that exist at both time step $k$ and $k+1$, and the density of newborn trajectories at time step $k+1$.

The backward smoothing equation \eqref{eq_smoothing}, in general, cannot be computed in closed form. An effective solution to evaluate the multi-trajectory posterior is given by backward simulation, which allows us to draw samples in the multi-object trajectory space \cite{xia2022multiple}. Performing backward simulation requires a backward kernel to recursively draw samples of sets of trajectories.

The backward kernel for sets of trajectories conditioned on the set ${\bf Y}$ of trajectories in the time interval $k+1:K$ and the sequence of measurement sets up to and including time step $K$, satisfies \cite[Lemma 3]{xia2022multiple}
\begin{equation}
    \label{eq_bs}
    \pi_{k:K|K}({\bf X}|{\bf Y}) \propto \pi_{k:k+1|k}({\bf X}_{k:k+1})\delta_{{\bf Y}}({\bf X}_{k+1:K}).
\end{equation}
That is, the backward kernel $\pi_{k:K|K}({\bf X}|{\bf Y})$ is proportional to the predicted multi-trajectory density $\pi_{k:k+1|k}({\bf X}_{k:k+1})$ if ${\bf Y} = {\bf X}_{k+1:K}$, and zero otherwise. Evaluating the backward kernel $\pi_{k:K|K}({\bf X}|{\bf Y})$ considers all different ways of associating the trajectories in ${\bf X}_{k:k+1}$ to the trajectories in ${\bf Y}$.

In this paper, the forward filtering is achieved using a PHD filter \cite{phdpoint}, and the PMB multi-object filtering densities $f_{k|k}(\mathbf{x})$ before the Poisson approximation are stored at each time step. Then backward simulation is performed on these PMB filtering densities to obtain the approximate multi-trajectory posterior.

\section{Hybrid PHD-PMB Trajectory Smoother}\label{sec_hybrid_phd_pmb}

\begin{figure}[!t]
    \centering
    \includegraphics[width=\columnwidth]{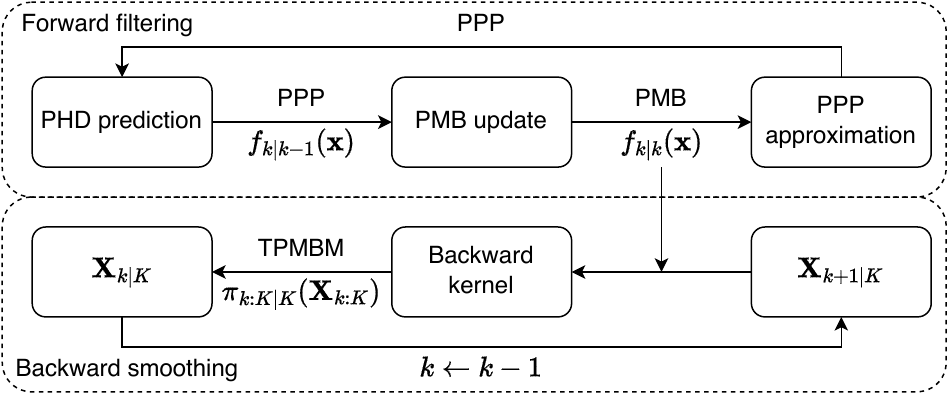}
    \caption{Forward filtering and backward smoothing recursions of hybrid PHD-PMB trajectory smoother. TPMBM represents trajectory PMBM \cite{granstrom2018poisson}.}
    \label{fig_diagram}
\end{figure}

In this section, we present the forward filtering and backward smoothing recursions of the hybrid PHD-PMB trajectory smoother. An illustration of the complete recursion is shown in Fig. \ref{fig_diagram}. We also discuss some practical considerations for a tractable linear Gaussian implementation.

\subsection{Forward PHD Filtering}

\begin{lemma}\label{prop_phd_predict}
    Given the Poisson multi-object filtering density $f_{k-1|k-1}(\mathbf{x})$ of the form \eqref{eq_ppp} with Poisson intensity $\lambda_{k-1|k-1}(\cdot)$ and the multi-object dynamic model described in Section \ref{sec_model}, the predicted multi-object density is Poisson, with intensity
    \begin{equation}\label{eq_predicted_ppp}
        \lambda_{k|k-1}(x) = \lambda^B(x) + \left\langle \lambda_{k-1|k-1} , g(x|\cdot) p^S(\cdot) \right\rangle.
    \end{equation}
\end{lemma}
Lemma \ref{prop_phd_predict} is given by the PHD prediction step \cite{phdpoint}.

\begin{lemma}\label{prop_phd_update}
    Given the Poisson multi-object predicted density $f_{k|k-1}(\mathbf{x})$ of the form \eqref{eq_ppp} with Poisson intensity $\lambda_{k|k-1}(\cdot)$ and the multi-object measurement model described in Section~ \ref{sec_model}, the updated multi-object density by the measurement set $\mathbf{z}_k = \{z_1,\dots,z_{m_k}\}$ is a PMB of the form \eqref{eq_pmb}, with $m_k$ Bernoulli components, one for each measurement. The updated Poisson intensity is 
    \begin{equation}\label{eq_pmb_ppp}
        \lambda_{k|k}^p(x) = \left(1-p^D(x)\right)\lambda_{k|k-1}(x),
    \end{equation}
    and the $i$-th new Bernoulli component created by measurement $z_k^i$, $i\in\{1,\dots,m_k\}$, is parameterized by
    \begin{subequations}\label{eq_pmb_ber}
        \begin{align}
            r_{k|k}^i &= \frac{\left\langle \lambda_{k|k-1},\ell\left(z_k^i | \cdot\right) p^D(\cdot)\right\rangle}{\lambda^C\left(z_k^i\right)+\left\langle \lambda_{k|k-1},\ell\left(z_k^i | \cdot\right) p^D(\cdot)\right\rangle}, \\
            p_{k|k}^i(x) &= \frac{\ell\left(z_k^i | x\right) p^D(x) \lambda_{k|k-1}(x)}{\left\langle \lambda_{k|k-1},\ell\left(z_k^i | \cdot\right) p^D(\cdot)\right\rangle}.\label{eq_ber_density}
        \end{align}
    \end{subequations}
\end{lemma}
Note that we use superscript $p$ in \eqref{eq_pmb_ppp} to represent the updated PPP intensity before the Kullback-Leibler divergence (KLD) minimization. Lemma \ref{prop_phd_update} is a special case of the PMBM update step \cite[Theorem 2]{pmbmpoint} with the predicted MB mixture being empty. In the updated PMB, the PPP represents missed detected objects, whereas the MB represents detected objects.

\begin{lemma}\label{eq_ppp_approx}
    The best PPP approximation of the updated PMB density in terms of minimizing the KLD yields the Poisson multi-object density with intensity
    \begin{align}\label{eq_ppp_approx_}
        \lambda_{k|k}(x) &= \lambda^p_{k|k}(x) + \sum_{i=1}^{m_k} r^i_{k|k}p^i_{k|k}(x)\nonumber \\
        &= \left(1-p^D(x)\right) \lambda_{k|k-1}(x) + \nonumber \\ &~~~\sum_{i=1}^{m_k} \frac{\ell\left(z_k^i | x\right) p^D(x) \lambda_{k|k-1}(x)}{\lambda^C\left(z_k^i\right)+\left\langle \lambda_{k|k-1},\ell\left(z_k^i | \cdot\right) p^D(\cdot)\right\rangle}.
    \end{align}
\end{lemma}
Lemma \ref{eq_ppp_approx} is proved in \cite[Sec. IV]{recycle}, and it is the same as the PHD update step \cite{phdpoint}. Note that in PHD filtering, the multi-object intensity \eqref{eq_ppp_approx_} is recursively computed over time without explicitly computing the Bernoulli densities \eqref{eq_pmb_ber}. Lemma \ref{prop_phd_update} and Lemma \ref{eq_ppp_approx} describe the PHD filter update as consisting of two steps, where the first step is implicit and gives a PMB representation of the multi-object posterior as an intermediate result before PPP approximation. Lemma~\ref{eq_ppp_approx} can be also interpreted as recycling all the Bernoulli components in PMB filtering \cite{recycle}, meaning that every Bernoulli component is implicitly approximated as a PPP. This yields a less accurate cardinality distribution, but it greatly simplifies the subsequent update steps as the approximate multi-object posterior only contains a single global hypothesis. Since the PMB is a more accurate representation of the multi-object posterior than the PPP, performing backward smoothing on the PMB filtering densities before PPP approximation allows us to exploit more information freely available in forward filtering.  

\subsection{Backward PMB Smoothing}

We first present the backward kernel \eqref{eq_bs} for PMB filtering densities. Then, we describe how to recursively draw samples of sets of trajectories from the backward kernel.
\begin{theorem}
    \label{lemma_pmb_smoothing}
    Given the PMB filtering density $f_{k|k}(\cdot)$, parameterized by Poisson intensity \eqref{eq_pmb_ppp} and Bernoulli densities \eqref{eq_pmb_ber}, at time step $k$, the set ${\bf Y}$ of trajectories in the time interval $k+1:K$, and the multi-trajectory dynamic model described in Section \ref{sec_tra_predict_model}, the backward kernel in the time interval $k:K$ \eqref{eq_bs} is a PMBM of the form
    
    \small
    \begin{align}
      \pi_{k:K|K}({\bf X}|{\bf Y}) &= \sum_{{\bf W} \uplus {\bf V} = {\bf X}} \pi_{k:K|K}^p({\bf W})\pi_{k:K|K}^{mb}({\bf V}|{\bf Y}),\label{eq_pmbm} \\
      \pi_{k:K|K}^p({\bf X}) &= e^{-\left\langle \lambda^p_{k:K|K},1 \right\rangle} \prod_{X \in {\bf X}}\lambda^p_{k:K|K}(X),\label{eq_ppp_bs} \\
      \pi_{k:K|K}^{mb}({\bf X}|{\bf Y}) &= \sum_{a\in {\cal A}_{k:K|K}}w^a \sum_{\biguplus^{n_{k:K|K}}_{l=1} {\bf X}^l = {\bf X}} \prod_{i=1}^{n_{k:K|K}} \pi^{i,a^i}_{k:K|K}\left({\bf X}^i\right).\label{eq_mbm}
    \end{align}
    \normalsize
    The set of global hypotheses is defined as
    \begin{multline}
        \label{eq_global_hypo}
        {\cal A}_{k:K|K} = \left\{ \left(a^1,\dots,a^{n_{k:K|K}} \right) : a^i \in \left\{1,\dots,h^i_{k:K|K}\right\}\forall i,\right.\\ \left. \left|{\cal M}^{i,a^i}_{k+1:K}\right| \leq 1, \biguplus_{i=1}^{n_{k:K|K}}{\cal M}^{i,a^i}_{k+1:K} = {\cal M}_{k+1:K}\right\},
    \end{multline}
    and the weight of global hypothesis $a\in{\cal A}_{k:K|K}$ satisfies 
    \begin{equation}
      \label{eq_glo_hyo_weight}
      w^a \propto \prod_{i=1}^{n_{k:K|K}}w^{i,a^i}_{k:K|K},
    \end{equation}
    where the proportionality means that normalization is required to ensure that $\sum_{a\in{\cal A}_{k:K|K}}w^a = 1$.

    In \eqref{eq_global_hypo}, ${\cal M}_{k+1:K}= \{1,\dots,n_{k+1:K}\}$ is the set of indices of trajectories in ${\bf Y} = \{Y^1,\dots,Y^{n_{k+1:K}}\}$ with $Y^1,\dots,Y^l$ being trajectories of objects that existed at time step $k+1$, and $Y^l,\dots,Y^{n_{k+1:K}}$ being trajectories of objects that appeared after time step $k+1$. Each trajectory in set ${\bf Y}$ creates a unique trajectory Bernoulli component, and thus the number of trajectory Bernoulli components in \eqref{eq_mbm} is $n_{k:K|K} = m_k + n_{k+1:K}$, indexed by variable $i\in\{1,\dots,n_{k:K|K}\}$. A global hypothesis is $a = (a^1,\dots,a^{n_{k:K|K}})$, where $a^i\in\{1,\dots,h^i_{k:K|K}\}$ is the index to the local hypothesis for the $i$-th trajectory Bernoulli component, and $h^i_{k:K|K}$ is its number of local hypotheses.
    
    The Poisson intensity in \eqref{eq_ppp_bs} is 
    \begin{equation}
      \lambda^p_{k:K|K}\left(t,x^{1:\nu}\right) = \delta_{k}[t]\delta_{1}[\nu]\left(1-p^S\left(x^1\right)\right)\lambda^p_{k|k}\left(x^1\right).
    \end{equation}
    For each trajectory Bernoulli component $i\in\{1,\dots,m_{k}\}$ in the predicted trajectory PMB $\pi_{k:k+1|k}({\bf X}_{k:k+1})$, there are $h_{k:K|K}^i = l+1$ local hypotheses. The local hypothesis that corresponds to the case that the trajectory ended at time step $k$, is given by ${\cal M}_{k+1:K}^{i,1} = \emptyset$ and 
    \begin{subequations}
      \begin{align}
        w^{i,1}_{k:K|K} &= 1-r^{i}_{k|k}+r^i_{k|k}\left\langle p^i_{k|k},1-p^S \right\rangle,\\
        r^{i,1}_{k:K|K} &= \frac{r^i_{k|k}\left\langle p^i_{k|k},1-p^S \right\rangle}{w^{i,1}_{k:K|K}},\\
        p^{i,1}_{k:K|K}\left(t,x^{1:\nu}\right) &= \delta_k[t]\delta_{1}[\nu]\frac{p^i_{k|k}\left(x^1\right)\left(1-p^S\left(x^1\right)\right)}{\left\langle p^i_{k|k},1-p^S \right\rangle}.
      \end{align}
    \end{subequations}
    
    The local hypothesis that corresponds to the case that the trajectory Bernoulli component is updated by trajectory $Y^j = (t^j,y^{1:\nu^j})$, $j\in\{1,\dots,l\}$ (present at time step $k+1$, i.e., $t^j=k+1$), is given by ${\cal M}_{k+1:K}^{i,j+1} = \{j\}$ and
    \begin{subequations}
      \begin{align}
        w^{i,j+1}_{k:K|K} &= r^i_{k|k}\left\langle p^i_{k|k},g\left(y^1|\cdot\right)p^S \right\rangle,\\
        r^{i,j+1}_{k:K|K} &= 1,\\
        p^{i,j+1}_{k:K|K}\left(t,x^{1:\nu}\right) &=\delta_{k}[t]\delta_{\nu^j+1}[\nu]\delta_{y^{1:\nu^j}}\left(x^{2:\nu}\right)\nonumber\\ &~~~\times\frac{g\left(y^1|x^1\right)p^i_{k|k}\left(x^1\right)p^S\left(x^1\right)}{\left\langle p^i_{k|k},g\left(y^1|\cdot\right)p^S \right\rangle}.
      \end{align}
    \end{subequations}
  
    The trajectory Bernoulli component created by trajectory $Y^j = (t^j,y^{1:\nu^j})$, $j\in\{1,\dots,l\}$, 
    has two local hypotheses $h^i_{k:K|K} = 2$, $i = m_k + j$. The first local hypothesis represents the case that the Bernoulli comment does not exist, and it is given by ${\cal M}_{k+1:K}^{i,1} = \emptyset$ and 
    \begin{equation}
      \label{eq_non_valid}
      w^{i,1}_{k:K|K} = 1,\quad r_{k:K|K}^{i,1} = 0.
    \end{equation}
    The second one is given by ${\cal M}_{k+1:K}^{i,2} = \{j\}$ and
    \begin{subequations}
      \label{eq_newly_detected}
      \begin{align}
        w^{i,2}_{k:K|K} &= \lambda^B_{k+1}\left(y^1\right) + \left\langle \lambda^p_{k|k},g\left(y^1|\cdot\right)p^S \right\rangle,\\
        r^{i,2}_{k:K|K} &= 1,\\
        p^{i,2}_{k:K|K}\left(t,x^{1:\nu}\right) &= \underline{w}^{i,2}_{k:K|K}\delta_{\left(t^j,y^{1:\nu^j}\right)}\left(t,x^{1:\nu}\right)\nonumber\\
        &~~~+\overline{w}^{i,2}_{k:K|K}\delta_{k}[t]\delta_{\nu^j+1}[\nu]\delta_{y^{1:\nu^j}}\left(x^{2:\nu}\right)\nonumber \\
        &~~~\times \frac{g\left(y^1|x^1\right)\lambda^p_{k|k}\left(x^1\right)p^S\left(x^1\right)}{\left\langle \lambda^p_{k|k},g\left(y^1|\cdot\right)p^S \right\rangle},\\
        \underline{w}^{i,2}_{k:K|K} &= \frac{\lambda^B_{k+1}\left(y^1\right)}{w^{i,2}_{k:K|K}},\\
        \overline{w}^{i,2}_{k:K|K} &= 1- \underline{w}^{i,2}_{k:K|K}.
      \end{align}
    \end{subequations}
  
    Finally, the trajectory Bernoulli component created by trajectory $Y^j$, $j\in\{l+1,\dots,n_{k+1:K}\}$ (not present at time step $k+1$) has a single local hypothesis $h_{k:K|K}^i = 1$, $i = m_k + j$, given by ${\cal M}_{k+1:K}^{i,1} = \{j\}$ and
    \begin{align}
        w^{i,1}_{k:K|K} &= 1, ~ r^{i,1}_{k:K|K} = 1, ~
        p^{i,1}_{k:K|K}(X) = \delta_{Y^j}(X).
    \end{align}
\end{theorem}
Theorem \ref{lemma_pmb_smoothing} is proved in \cite[App. E]{xia2022multiple}. 

% \begin{remark}
%     The predicted trajectory PPP in $\pi_{k:k+1|k}({\bf X}_{k:k+1})$ has different interpretations in \cite{xia2022multiple} and in this work. In \cite{xia2022multiple}, the PPP represents trajectories that have never been detected, whereas in this work it represents missed detected trajectories due to the PPP approximation after the PHD update step. This difference further leads to different interpretations of the local hypothesis that a trajectory is associated to the trajectory PPP \eqref{eq_newly_detected}. In \cite{xia2022multiple}, the second mixture component in \eqref{eq_newly_detected} represents the case that the trajectory was first detected at time step $k+1$, but in this work it represents the case that the trajectory might be first detected at time step $k+1$, or at any time before.
% \end{remark}

By running the backward simulation for sets of trajectories $T$ times for $k = K-1,\dots,1$, where we recursively draw samples of ${\bf X}_{k:K}$ from the backward kernel \eqref{eq_bs}, we can obtain $T$ samples of $\{{\bf X}^{(i)}_{1:K}\}_{i=1}^T$. This gives a particle representation of the multi-trajectory density
\begin{equation}
  \label{eq_particle}
  \pi({\bf X}_{k:K}) \approx \sum_{i=1}^{T}\frac{1}{T}\delta_{{\bf X}^{(i)}}({\bf X}_{k:K}),
\end{equation}
where the $i$-th particle has state ${\bf X}^{(i)}$ and weight $1/T$. 

\subsection{Practical Considerations for a Tractable Implementation}  

In this work, we consider the linear Gaussian implementations of the hybrid PHD-PMB trajectory smoother. The linear Gaussian implementations of the PHD filter and the PMB smoother using backward simulation have been described in \cite{phdpoint2} and \cite{xia2022multiple}, respectively. Here, we highlight some practical considerations that facilitate an efficient implementation.

In forward PHD filtering, the Poisson intensity \eqref{eq_ppp_approx_} is of the form Gaussian mixture, and its number of mixture components increases with the number of measurements over time without bound. To obtain a tractable implementation, Gaussian mixture reduction needs to be performed by pruning Gaussian components with small weights and merging similar Gaussian components \cite{phdpoint2}.

We extract the PMB filtering densities after the PHD update step before PPP approximation. It should be noted that doing so only involves minor modifications of an existing implementation of the PHD update step. By comparing \eqref{eq_pmb_ber} and \eqref{eq_ppp_approx_}, we can see that the Poisson intensity in PMB is directly given by the first term in \eqref{eq_ppp_approx_}, and that the single-object density of the $i$-th Bernoulli density can be easily obtained by normalizing the $i$-th term in the summation in \eqref{eq_ppp_approx_}, where the normalization factor gives the existence probability $r_{k|k}^i$ in \eqref{eq_pmb_ber}. In addition, since the predicted PPP intensity \eqref{eq_predicted_ppp} is a Gaussian mixture, the single-object density \eqref{eq_ber_density} of each Bernoulli component in the PMB is also a Gaussian mixture. This means that in backward simulation we may need many particles to find the mode of the multi-trajectory density $\pi_{1:K|K}(\mathbf{X})$. For efficient sampling, we can reduce each single-object density \eqref{eq_ber_density} to a single Gaussian by moment matching. 

In backward simulation, it is trivial to parallize the sampling process  since multiple backward sets of trajectories can be generated independently. The main challenge lies in sampling the MB mixture \eqref{eq_mbm} in the PMBM backward kernel \eqref{eq_pmbm}, which has an intractable number of MB components due to the unknown associations between the different components in the PMB filtering density and the trajectories. To avoid enumerating every global hypothesis, which is of combinatorial complexity, we can draw samples only from MB components with non-negligible weights \cite{xia2022multiple}. One way to do so is by solving a ranked assignment problem using Murty's algorithm \cite{crouse2016implementing}. In addition, ellipsoidal gating can be applied to remove unlikely local hypotheses to simplify the computation of the assignment problem.

\section{Simulation Results}

In this section, we evaluate the performance of the proposed hybrid PHD-PMB trajectory smoother in a simulation study in comparison with the PHD filter \cite{phdpoint2} and the trajectory PHD filter \cite{garcia2019trajectory}. 

% We also analyze the premature object death problem \cite{nagappa2011fast} in the hybrid PHD-PMB trajectory smoother via a toy example.

\begin{figure}[!t]
    \centering
    \includegraphics[width=\columnwidth]{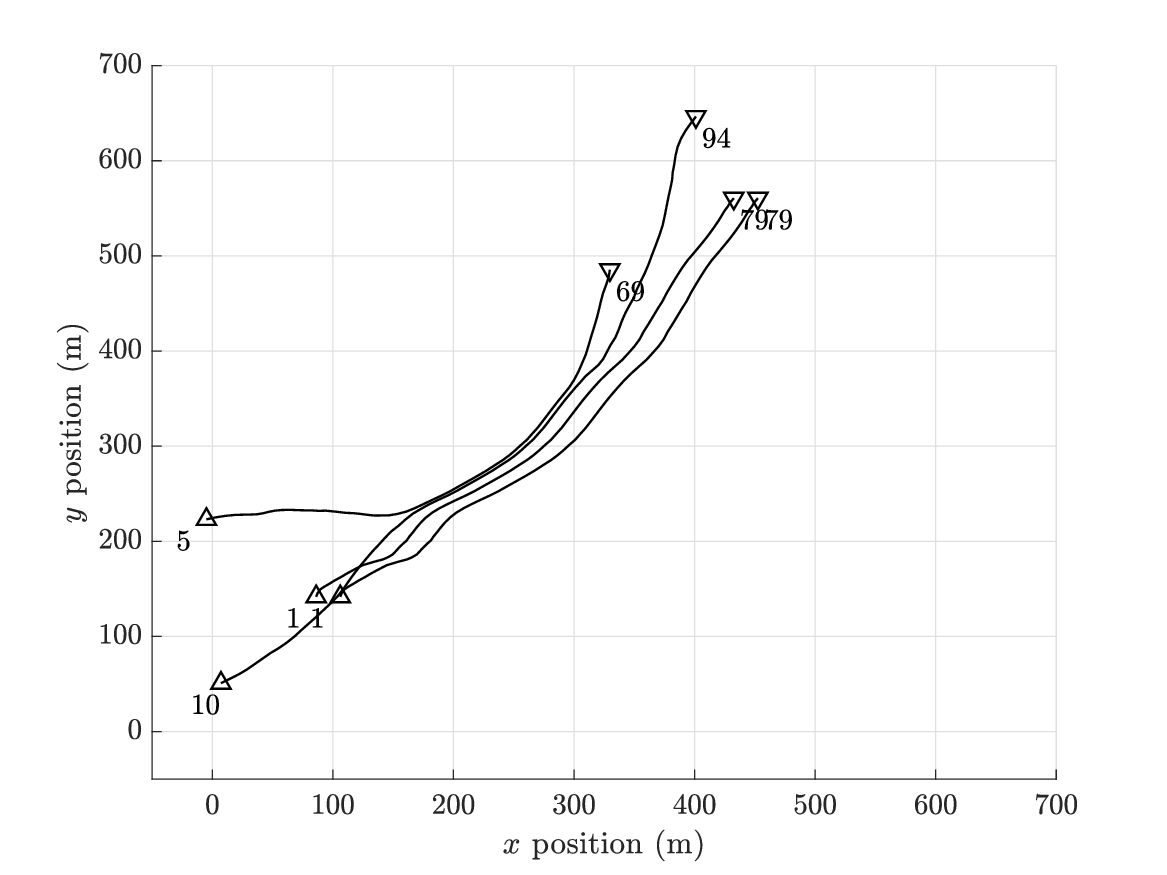}
    \caption{True object trajectories. The up triangles and the numbers next to them indicate the starting position and time steps of different trajectories. The down triangles and the numbers next to them indicate the ending position and time steps of different trajectories, respectively.}
    \label{fig_gt}
\end{figure}

We consider the same two-dimensional scenario as in \cite{garcia2019trajectory} with 100 time steps and 4 objects. The single object state is $x = [p_x,\dot{p_x},p_y,\dot{p_y}]^T$, consisting of position and velocity. The single object dynamic model is nearly constant velocity with transition matrix $F$ and process noise covariance $Q$ given by 
\begin{equation*}
  F = I_2 \otimes \begin{bmatrix}
    1 & T_s\\
    0 & 1
  \end{bmatrix},\quad Q = \sigma_q^2 I_2 \otimes \begin{bmatrix}
    T_s^3/3 & T_s^2/2\\
    T_s^2/2 & T_s
  \end{bmatrix},
\end{equation*}
where $I_2$ is a $2 \times 2$ identity matrix, $\otimes$ denotes the Kronecker product, $T_s = 0.5 \,\text{s}$ is the sampling period, and $\sigma_q = 1.8$. Each object survives with probability $p^S = 0.99$. The birth model is a PPP, and its Poisson intensity is a Gaussian mixture with three components. Each component has the same weight $0.1$ and covariance $\text{diag}([225, 100, 225, 100])$; the means are $[85,0,140,0]^T$, $[-5,0,220,0]^T$, and $[7,0,50,0]^T$. The ground truth object trajectories are illustrated in Fig. \ref{fig_gt}.

The single object measurement model is linear and Gaussian with observation matrix $H$ and measurement noise covariance $R$ given by
\begin{equation*}
  H = I_2 \otimes \begin{bmatrix}
    1 & 0
  \end{bmatrix},\quad R = \sigma_r^2 I_2,
\end{equation*}
where $\sigma_r = 2$. The detection probability is $p^D = 0.9$. The clutter is uniformly distributed in the area $[0\,\text{m},2000\,\text{m}]\times[0\,\text{m},2000\,\text{m}]$ with Poisson clutter rate $\gamma^C = 50$.

\begin{figure}[!t]
  \centering
  \includegraphics[width=\columnwidth]{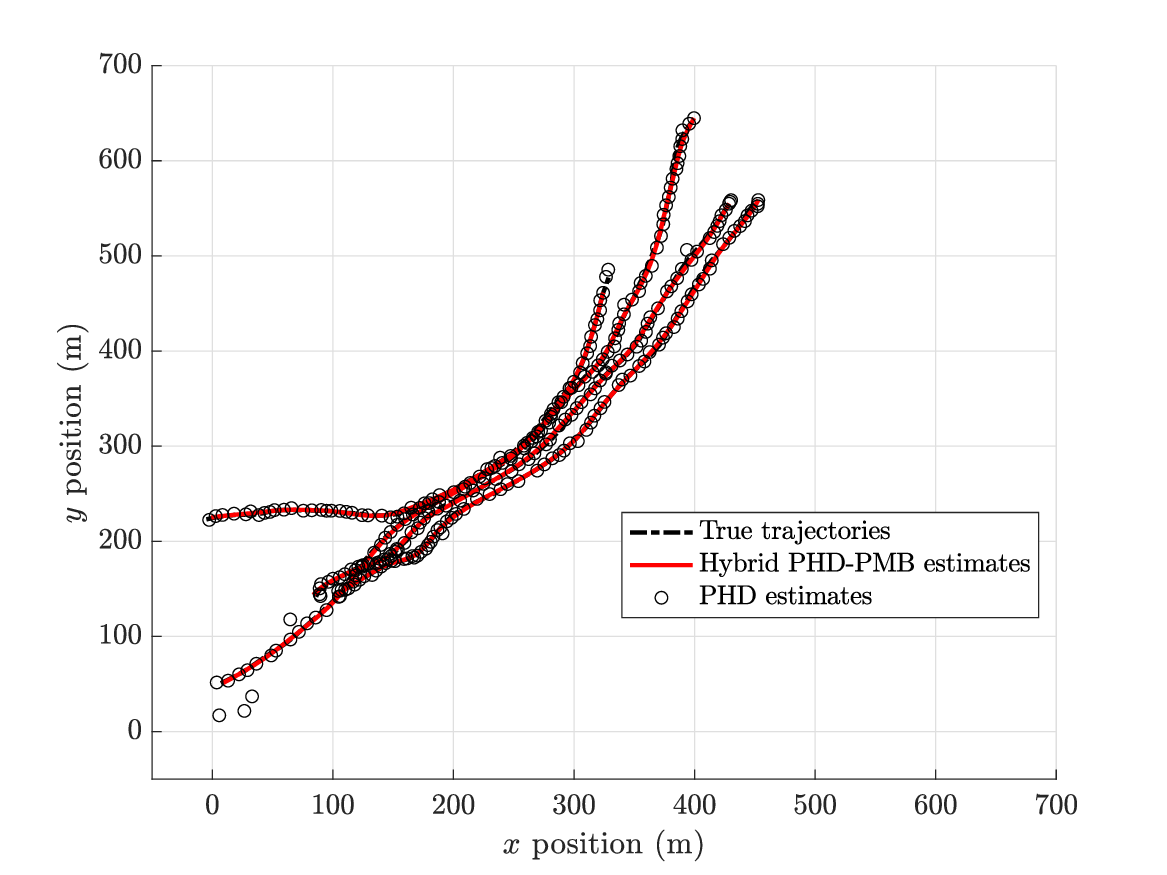}
  \caption{Exemplar output of the hybrid PHD-PMB trajectory smoother. It can be seen that the hybrid PHD-PMB trajectory smoother can produce smoothed estimates for all the trajectories. In addition, it can recover missed detections and disregard false detections in the forward PHD filtering.}
  \label{fig_output}
\end{figure}

The PHD filter \cite{phdpoint2} is implemented with pruning threshold $10^{-4}$, merging threshold $4$, and maximum number of Gaussian components $30$. Object state estimates are obtained using the estimator described in \cite[Sec.~9.5.4.4]{mahler2007statistical}. The trajectory PHD filter \cite{garcia2019trajectory} is implemented also with pruning threshold $10^{-4}$, absorption threshold $4$, and maximum number of Gaussian components $30$, and it performs smoothing-while-filtering without $L$-scan approximation. Its trajectory estimates are obtained using the estimator described in \cite[Sec. VII-D]{garcia2019trajectory}. The hybrid PHD-PMB trajectory smoother is implemented with $1000$ particles, and only a maximum of 100 global hypotheses are sampled. Its trajectory estimates are obtained using the estimator described in \cite[App. G]{xia2022multiple}. An exemplar output of the hybrid PHD-PMB trajectory smoother is shown in Fig. \ref{fig_output}.

The multi-object state estimation performance is evaluated using the generalized optimal subpattern assignment (GOSPA) metric \cite{gospa} at each time step with $\alpha = 2$, $c=10$, and $p=1$. GOSPA allows for the decomposition of the estimation error into localization error, missed and false detection errors. The multi-trajectory smoothing estimation performance is evaluated for the set of all trajectories using the trajectory GOSPA (TGOSPA) metric \cite{garcia2020metric} at the last time step with $c=10$, $p=1$, and $\gamma=1$. TGOSPA allows for the decomposition of the estimation error into localization error, missed and false detection errors, and track switch error. We note that the trajectory PHD filter cannot estimate the set of all trajectories accurately. To report the set of all trajectory estimates, we tag each Gaussian component and extract the trajectory estimates with unique tags via post-processing. We report the results averaged over 100 Monte Carlo runs.

\begin{figure*}[!t]
    \centering
    \includegraphics[width=0.49\columnwidth]{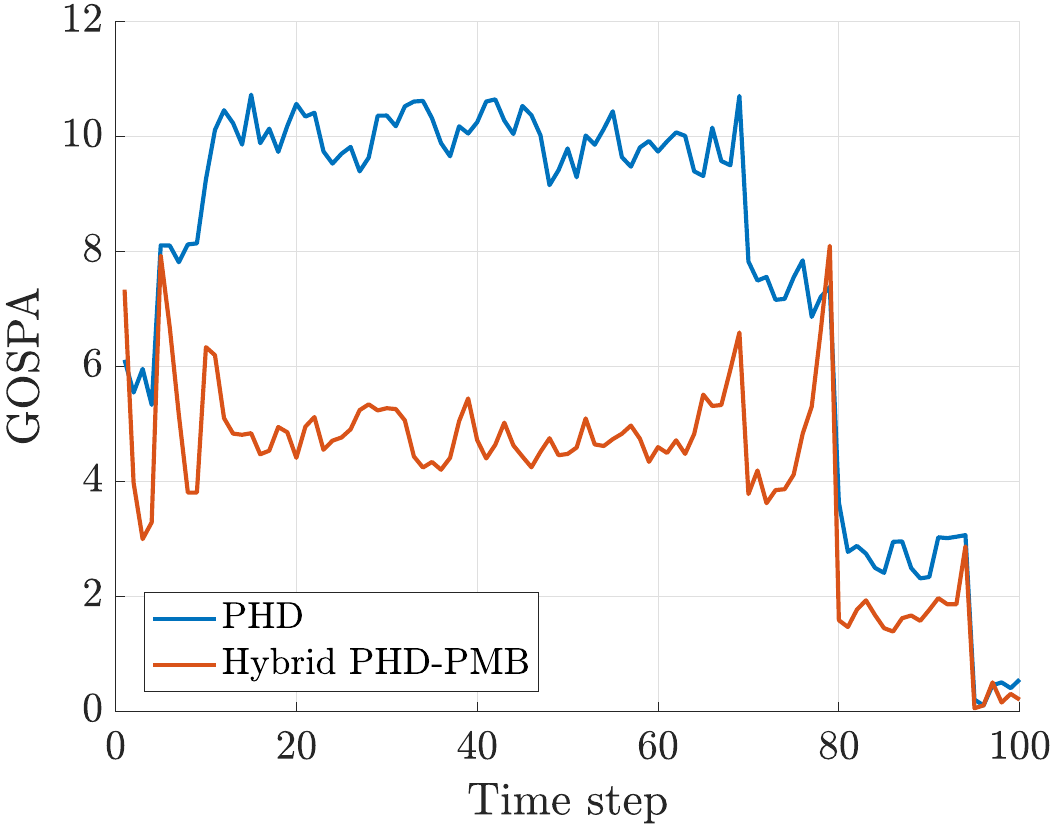}
    \includegraphics[width=0.49\columnwidth]{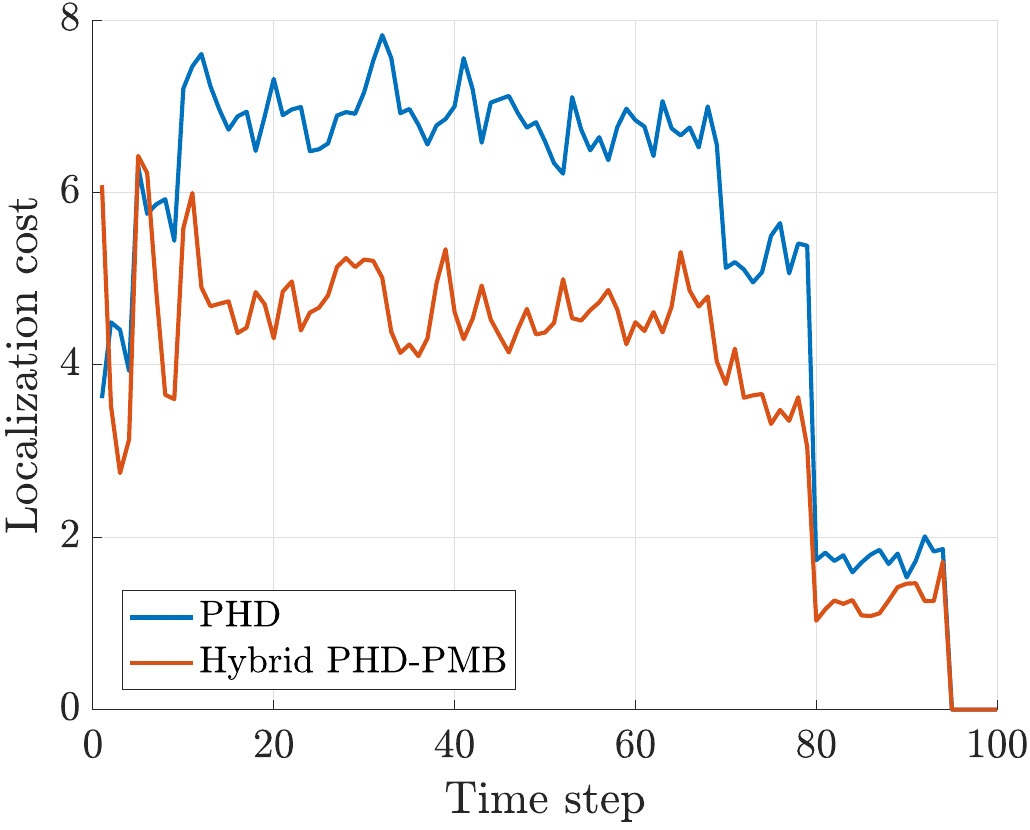}
    \includegraphics[width=0.49\columnwidth]{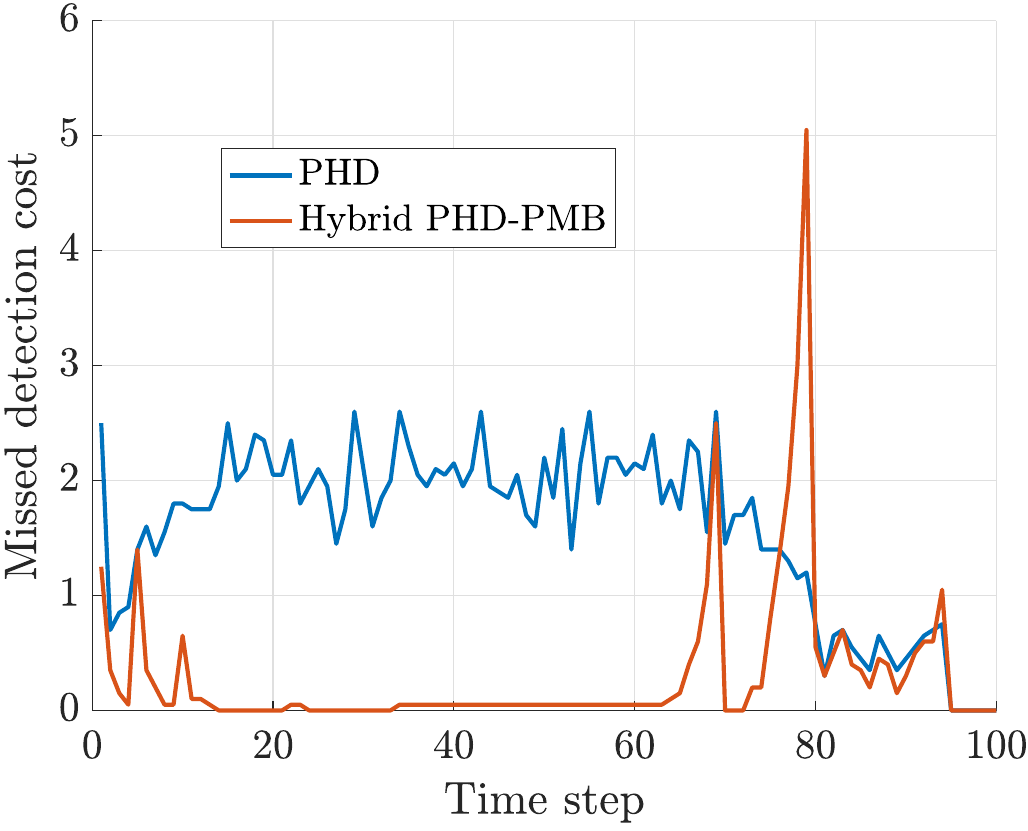}
    \includegraphics[width=0.49\columnwidth]{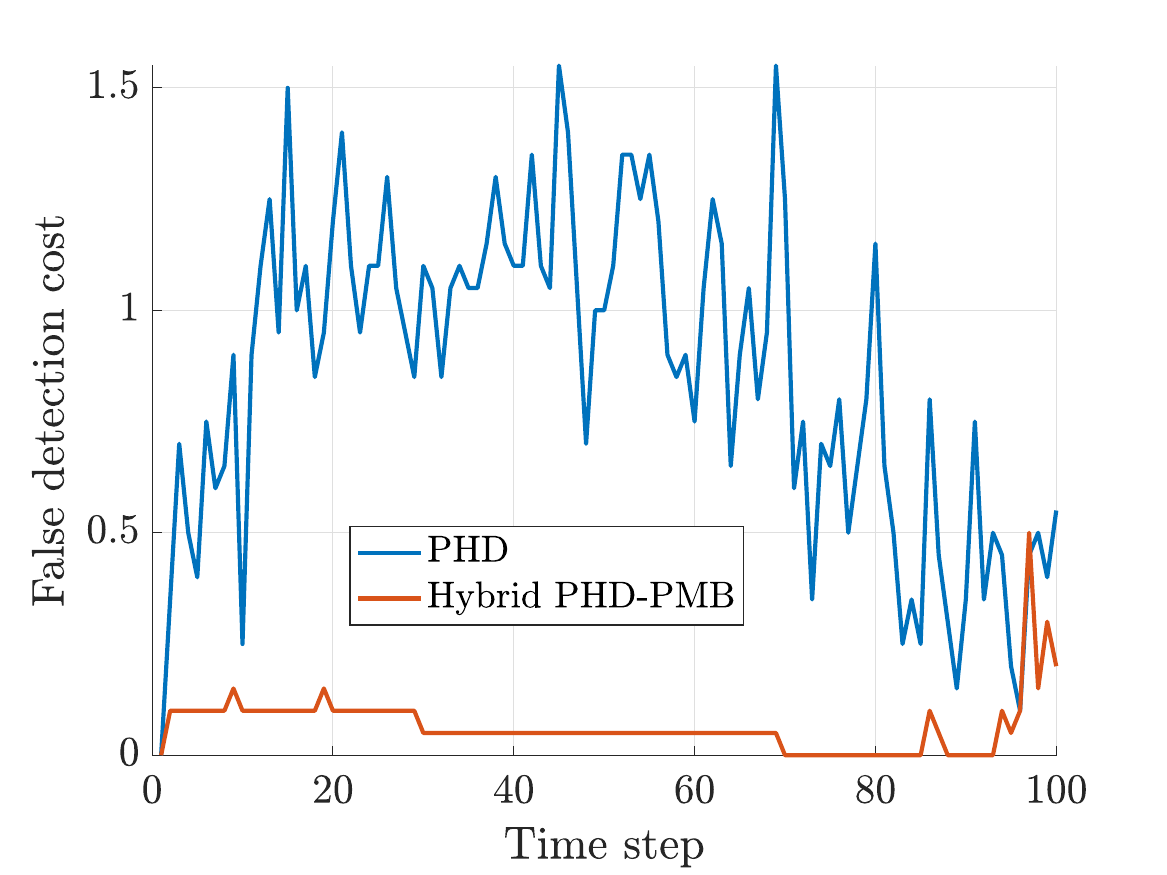}
    \caption{GOSPA error and its decomposition over time.}
    \label{fig_gospa}
    \vspace{-2mm}
\end{figure*}

\begin{table*}[!t]
  \centering
  \caption{Average GOSPA error and its decomposition per time step for different scenario parameters.}
  \begin{tabular}{c|cccc|cccc}
  \hline
  & \multicolumn{4}{c|}{PHD}                                  & \multicolumn{4}{c}{Hybrid PHD-PMB}                             \\ \hline
  & GOSPA & Localization & Missed & False & GOSPA      & Localization & Missed & False \\ \hline
  No change        & 7.82  & 5.38         & 1.58       & 0.86   & \underline{4.20} & 3.79      & 0.31             & 0.10            \\
  $\gamma^C = 10$  & 7.26  & 5.42         & 1.55       & 0.28    & \underline{4.06} & 3.76         & 0.23             & 0.06            \\
  $\gamma^C = 100$ & 8.28  & 5.41         & 1.59       & 1.27    & \underline{4.30} & 3.81         & 0.37             & 0.12            \\
  $p^D = 0.98$    & 6.53  & 5.73         & 0.35        & 0.45     & \underline{3.80} & 3.58         & 0.16             & 0.06            \\
  $p^D = 0.8$     & 10.95 & 5.21         & 2.55        & 3.19     & \underline{4.76} & 4.11         & 0.56             & 0.10    \\
  \hline       
  \end{tabular}
  \label{tab_gospa}
\end{table*}

\begin{figure*}[!t]
  \centering
  \includegraphics[width=0.19\textwidth]{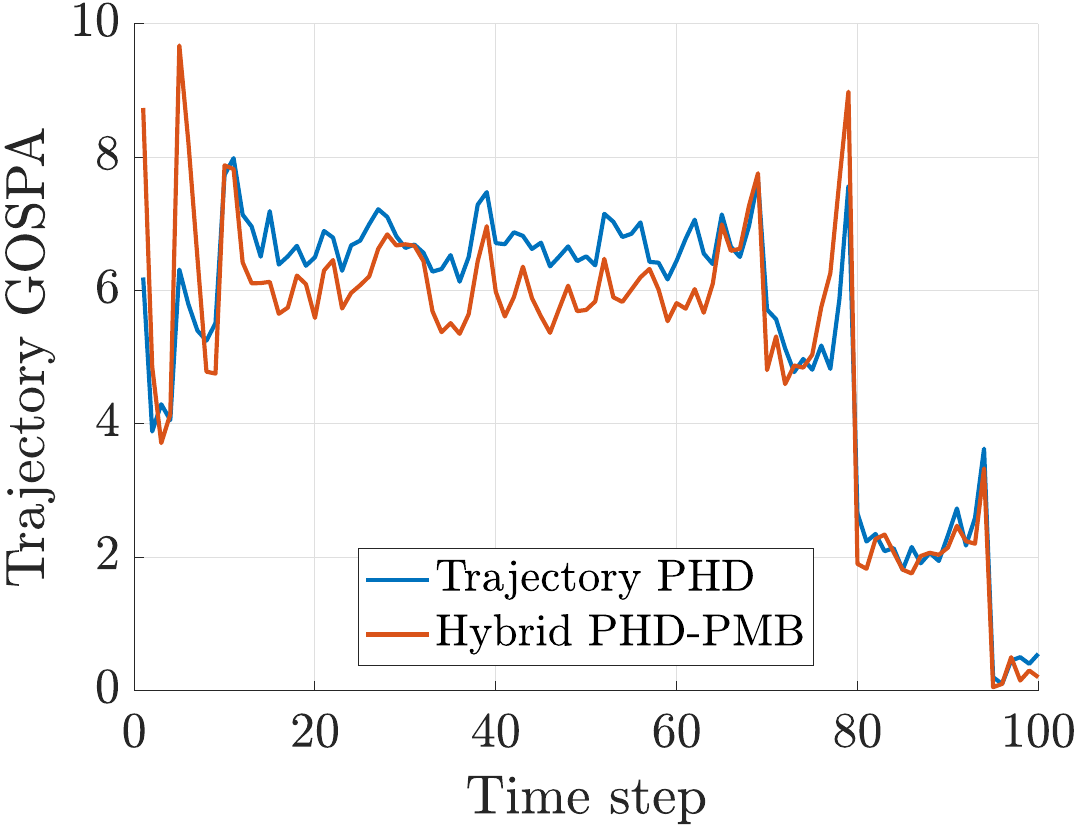}
  \includegraphics[width=0.19\textwidth]{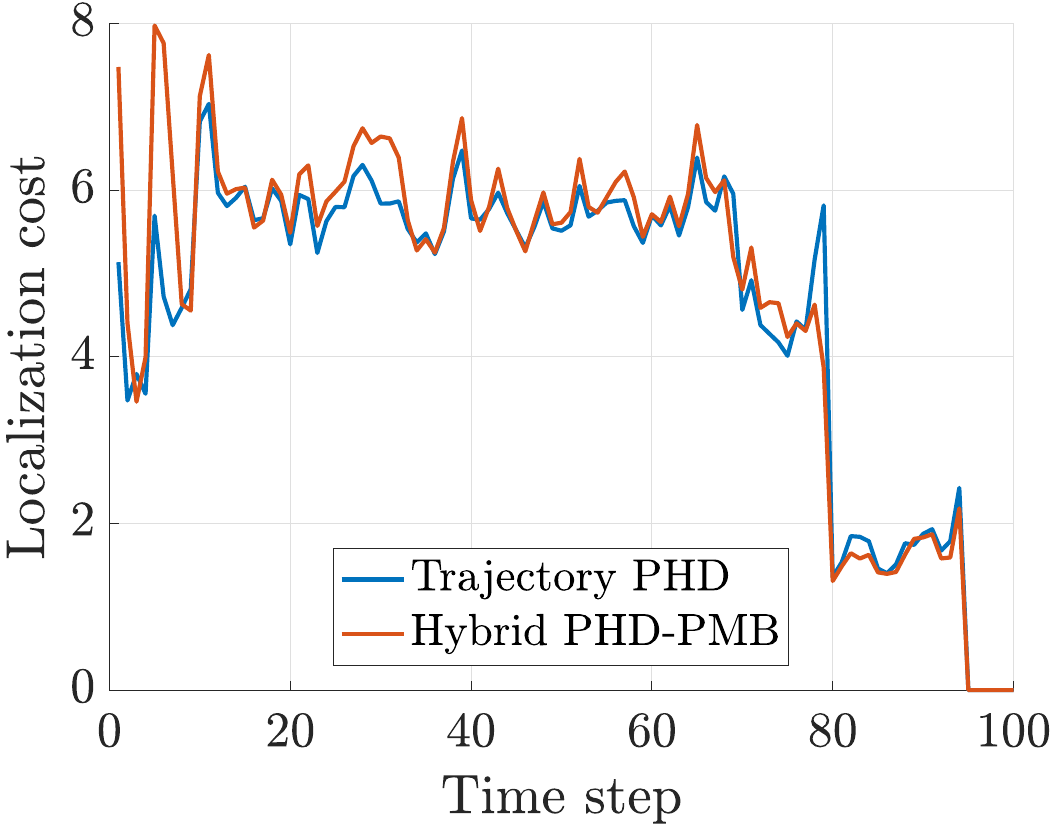}
  \includegraphics[width=0.19\textwidth]{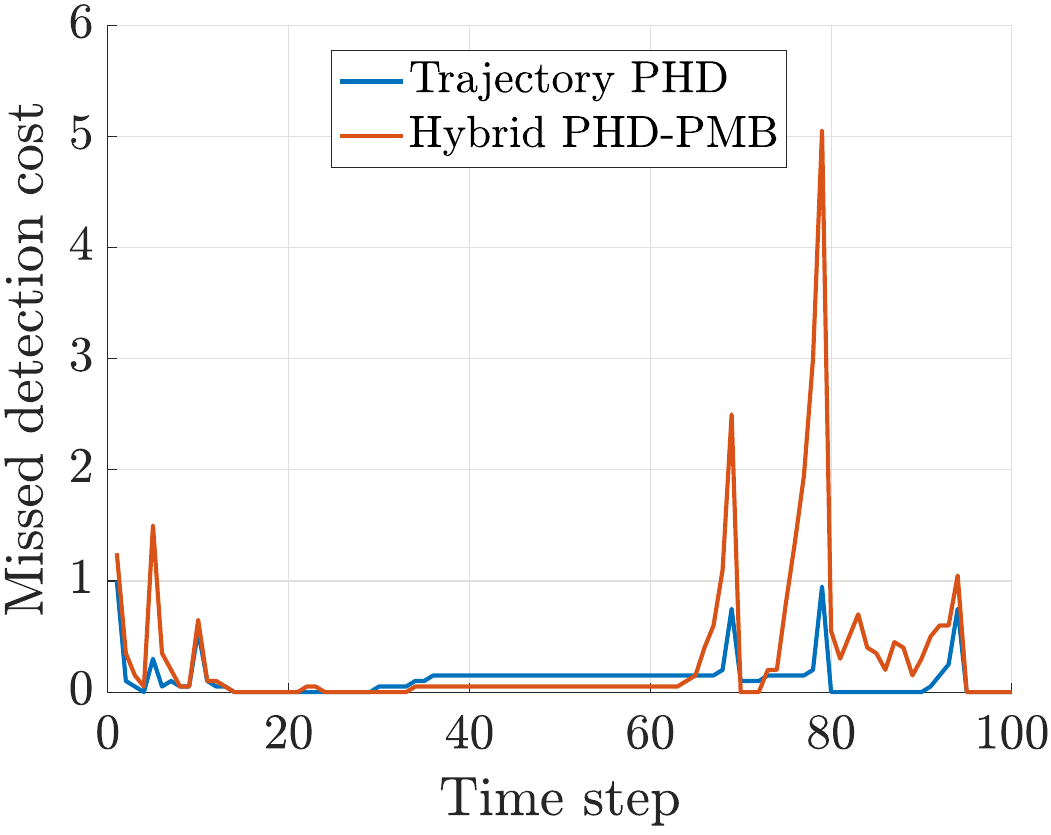}
  \includegraphics[width=0.19\textwidth]{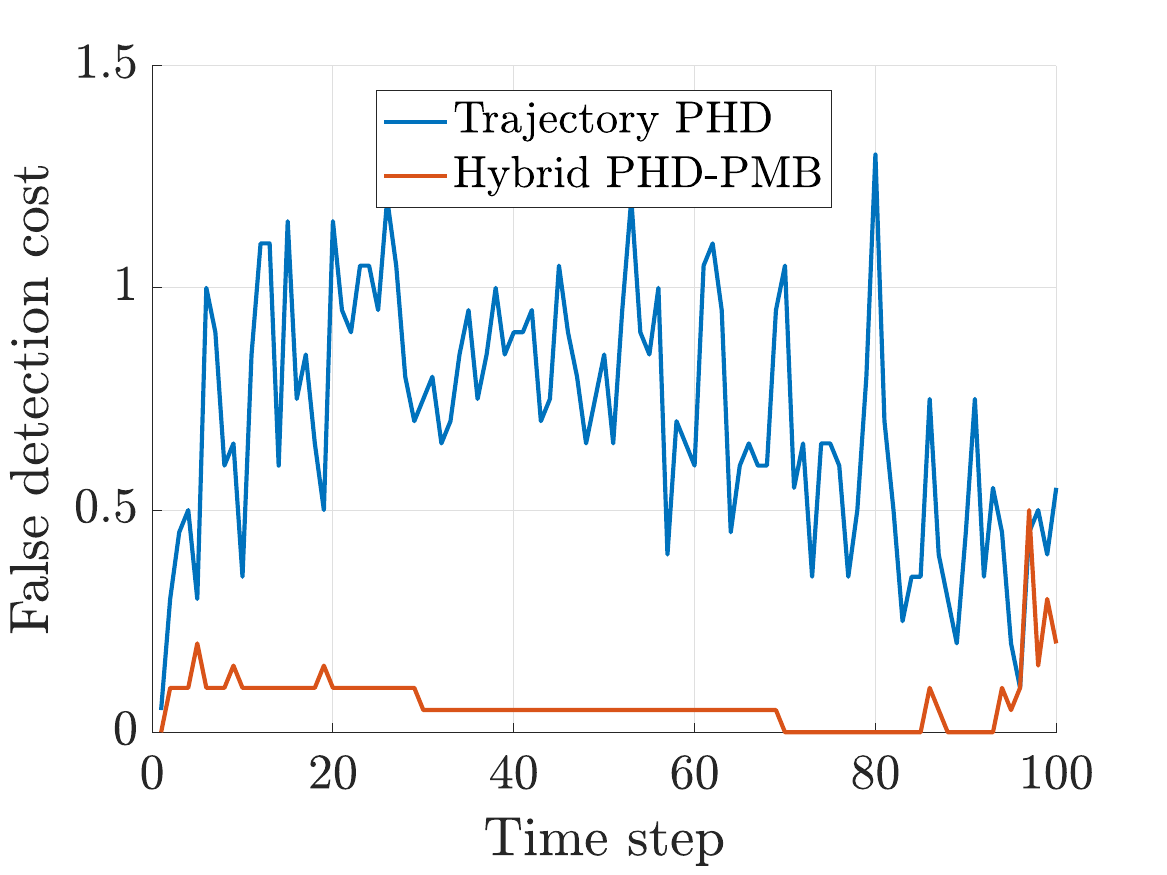}
  \includegraphics[width=0.19\textwidth]{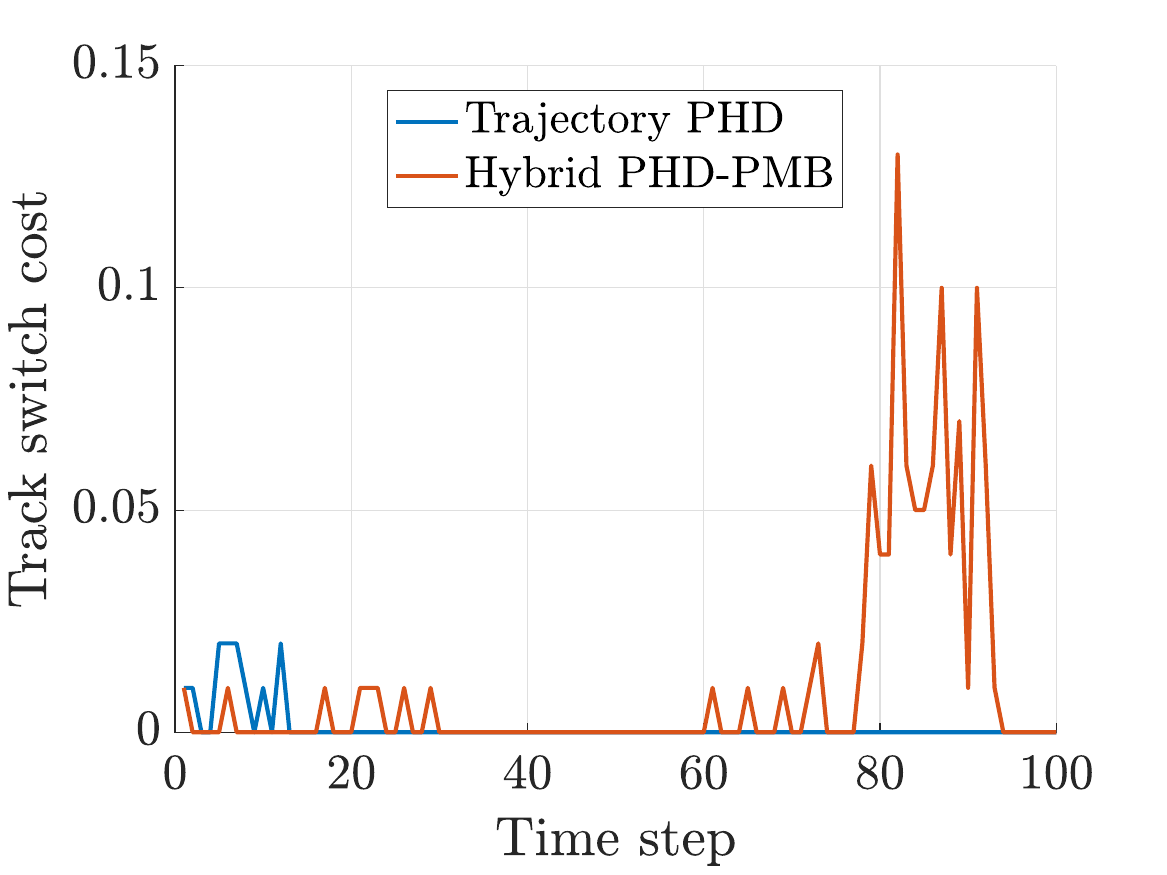}
  \caption{TGOSPA error and its decomposition over time.}
  \label{fig_tgospa}
  \vspace{-2mm}
\end{figure*}

\begin{table*}[!t]
  \centering
  \caption{Average TGOSPA error and its decomposition per time step for different scenario parameters.}
  \begin{tabular}{c|ccccc|ccccc}
  \hline
  & \multicolumn{5}{c|}{Trajectory PHD}                                                 & \multicolumn{5}{c}{Hybrid PHD-PMB}                                                  \\ \hline
  & TGOSPA & Localization & Missed & False & Track switch & TGOSPA & Localization & Missed & False & Track switch \\ \hline
  No change        & 5.43             & 4.61         & 0.12             & 0.71            & 0.00         & \underline{5.22}       & 4.79         & 0.31             & 0.10            & 0.01         \\
  $\gamma^C = 10$  & \underline{5.01}       & 4.65         & 0.13             & 0.23            & 0.00         & 5.14       & 4.82         & 0.23             & 0.07            & 0.02         \\
  $\gamma^C = 100$ & 5.90             & 4.68         & 0.18             & 1.04            & 0.00         & \underline{5.34}       & 4.84         & 0.37             & 0.12            & 0.01         \\
  $p^D = 0.98$     & 4.92             & 4.51         & 0.01             & 0.39            & 0.00         & \underline{4.78}       & 4.56         & 0.16             & 0.06            & 0.00         \\
  $p^D = 0.8$      & 6.28             & 4.86         & 0.29             & 1.13            & 0.00         & \underline{5.91}       & 5.22         & 0.56             & 0.10            & 0.02    \\
  \hline    
  \end{tabular}
  \label{tab_tgospa}
\end{table*}

Let us first study the multi-object estimation performance. The GOSPA error and its decomposition for the the PHD filter and the hybrid PHD-PMB trajectory smoother over time are shown in Fig. \ref{fig_gospa}. Overall, the hybrid PHD-PMB trajectory smoother outperforms the PHD filter in terms of localization errors, missed and false detection errors by a significant margin. One exception is that the hybrid PHD-PMB trajectory smoother presents higher localization errors at the first few time steps due to the mismatch between the Gaussian means in the birth intensity and the true object birth positions. Moreover, the PHD-PMB trajectory smoother shows one apparent anomaly that it has increased missed detection error a few time steps before object disappearing. This anomaly is due to the premature object death, a problem also observed in PHD smoothers \cite{mahler2012forward,nagappa2011fast}, where missed detections happen at earlier time steps when objects disappear, with a lag equal to the smoothing lag. However, we note that in the hybrid PHD-PMB trajectory smoother this problem is less severe in the sense that it does not always happen for every disappearing object, and that missed detections due to premature object death are not propagated all the way back to earlier time steps. We also show the average GOSPA error and its decomposition per time step for different scenario parameters in Table \ref{tab_gospa}. We can see that the hybrid PHD-PMB trajectory smoother consistently outperforms the PHD filter by a significant margin, and it is especially robust to false detections.

We then discuss the trajectory estimation performance of the hybrid PHD-PMB trajectory smoother and the trajectory PHD filter. The TGOSPA error and its decomposition for the trajectory PHD filter and the hybrid PHD-PMB trajectory smoother over time are shown in Fig. \ref{fig_tgospa}. The hybrid PHD-PMB trajectory smoother surpasses the trajectory PHD filter by producing fewer false detections. However, the trajectory PHD filter exhibits slightly superior performance in localization accuracy, reduced missed detections, and fewer track switches. The average TGOSPA error and its decomposition per time step for different scenario parameters are presented in Table \ref{tab_tgospa}. We can see that the hybrid PHD-PMB trajectory smoother generally outperforms the trajectory PHD filter, except for the scenario with low clutter rate.

\section{Conclusions}
In this paper, we have presented a hybrid PHD-PMB multi-object smoother, which performs backward simulation using the PMB densities obtained in the update step of forward PHD filtering to provide smoothed trajectory estimates. The hybrid PHD-PMB trajectory smoother makes it possible for the PHD filter to generate smoothed trajectory estimates for all the objects. Possible future work includes the development of a hybrid PHD-PMB trajectory smoother for non-Gaussian, non-linear scenarios using sequential Monte Carlo implementations and experiments with real data \cite{liu2022gnn}.

\bibliographystyle{IEEEtran}
\bibliography{mybibli.bib}

\end{document}